\documentclass[sigconf]{acmart}

\usepackage{multirow}

\newcommand{\todo}[1]{}
\newcommand{\todoq}[1]{}
\newcommand{\citetodo}[1]{}

\copyrightyear{2026}
\acmYear{2026}
\setcopyright{cc}
\setcctype{by}
\acmConference[CIKM '26]{Proceedings of the 35th ACM International
  Conference on Information and Knowledge Management}{November 07--11,
  2026}{Rome, Italy}
\acmBooktitle{Proceedings of the 35th ACM International Conference on
  Information and Knowledge Management (CIKM '26), November 07--11,
  2026, Rome, Italy}
\acmDOI{10.1145/3799682.3840042}
\acmISBN{979-8-4007-2539-5/2026/11}

\begin{document}

\title{Enriching Sequential Recommendation with\texorpdfstring{\\}{ }Graph Laplacian Positional Embeddings}

\author{Ekaterina Trushkova}
\orcid{0009-0002-3398-5224}
\correspondingauthor
\email{eksetrushkova@edu.hse.ru}
\affiliation{%
  \institution{HSE University}
  \city{Moscow}
  \country{Russian Federation}
}

\author{Artur Gimranov}
\orcid{0009-0006-4296-3165}
\email{agimranov@hse.ru}
\affiliation{%
  \institution{HSE University}
  \city{Moscow}
  \country{Russian Federation}
}

\author{Anton Lysenko}
\orcid{0000-0002-9369-7104}
\email{av.lysenko@hse.ru}
\affiliation{%
  \institution{HSE University}
  \city{Moscow}
  \country{Russian Federation}
}

\renewcommand{\shortauthors}{Ekaterina Trushkova, Artur Gimranov, and Anton Lysenko}

\begin{abstract}

Sequential recommenders typically rely on learnable positional embeddings to encode the order of user interactions. In this work, we ask whether this ordinal signal can be replaced by a structural one derived from the item space. We propose to use Laplacian positional embeddings in SASRec: we build an item co-occurrence graph from training interactions, compute eigenvectors of its symmetric normalized Laplacian, and use them as frozen graph-derived positional embeddings. The backbone architecture and training objective remain unchanged. Experiments on four public sequential-recommendation benchmarks show that this simple replacement improves SASRec performance on most ranking metrics and remains competitive with strong positional and temporal encoding baselines. These findings indicate that item-item graph structure can be an effective substitute for standard ordinal positional embeddings in sequential recommendation.
\end{abstract}



\maketitle

\section{Introduction}
\label{sec:intro}

Sequential recommendation aims to predict the next item a user will interact with from their chronologically ordered history. Self-attention models, and SASRec \cite{sasrec2018} in particular, have become a widely adopted architecture for this task because they can model long-range dependencies within interaction sequences~\cite{bert4rec,petrov2022systematic}. In these models, positional embeddings play a central role: they provide the Transformer with information about the order of interactions, allowing it to distinguish the same items appearing at different positions in a sequence.

However, chronological order is not the only source of useful structure in recommendation data. User interactions also induce a global item--item structure: items that are frequently consumed by the same users are likely to be related, even if they do not appear in the same local sequential context. Graph-based recommenders explicitly exploit such structure by operating on user--item or item--item graphs~\cite{lightgcn2020,Wang2019-NGCF}. In contrast, standard sequential recommenders usually access this signal only indirectly through the training objective and item embeddings. This raises a natural question: can the item--item graph be injected into a sequential recommender through its positional encoding mechanism?

A straightforward way to combine graph and sequence information is to add a graph encoder to a sequential model. Several hybrid approaches follow this direction, but they introduce additional trainable components and couple graph representation learning with sequential training~\cite{Wang2022-MCLSR,Baikalov2024-MRGSRec}. In this work, we study a lighter alternative. Instead of modifying the SASRec architecture, we replace its learnable positional embeddings with graph-derived embeddings computed once before training. Specifically, we construct an item co-occurrence graph from the training interactions, compute eigenvectors of the symmetric normalized graph Laplacian, and use them as frozen Laplacian positional embeddings. Thus, the positional slot of SASRec is repurposed from encoding sequence indices to encoding structural positions of items in the global item graph.

This design is motivated by Laplacian positional encodings used in graph Transformers, where spectral coordinates provide a structural notion of node position~\cite{Dwivedi2020-GraphTransformer,Kreuzer2021-SAN,Rampasek2023-GraphGPS}. We adapt this idea to sequential recommendation: each item receives a spectral embedding determined by its location in the item co-occurrence graph, and this embedding is added to the item representation at the input layer. Importantly, the graph eigendecomposition is performed only once, the resulting embeddings are kept fixed, and the sequential backbone, loss, and training procedure remain unchanged.

We empirically evaluate the proposed approach on four public sequential-recommendation benchmarks. The results show that replacing standard learnable positional embeddings with Laplacian positional embeddings improves SASRec on most ranking metrics and is competitive with strong positional and temporal encoding baselines, including rotary positional embeddings and TiSASRec. These findings suggest that item--item graph structure provides recommendation signal that is not fully captured by ordinal positional embeddings alone.

Our contributions are as follows:
\begin{itemize}
    \item We propose a lightweight graph-derived positional encoding for SASRec based on eigenvectors of the normalized Laplacian of the item co-occurrence graph.
    \item We show how this spectral signal can be used as a frozen replacement for standard learnable positional embeddings without changing the SASRec backbone or training objective.
    \item We provide an empirical comparison against learnable, rotary, and time-aware positional encoding baselines, showing that graph-derived positional embeddings improve SASRec on most ranking metrics and remain competitive with stronger positional alternatives.
\end{itemize}

The code and data to reproduce our experiments are available at \url{https://github.com/trsh2310/lap_pe.git}.

\section{Related Work}
\label{sec:related}

\paragraph{Sequential recommendation.}
Self-attention based models have established the state of the art for
next-item prediction. SASRec applies the
Transformer encoder to user interaction sequences with a learnable
positional embedding and a sampled binary cross-entropy loss; BERT4Rec
\cite{bert4rec} adopts a bidirectional masked-item
objective. SASRec+ \cite{Klenitskiy2023-SASRecPlus} subsequently
showed that replacing SASRec's sampled loss with a full-catalog
cross-entropy substantially closes the gap to BERT4Rec while
preserving SASRec's lower training cost. We adopt SASRec (with the full-catalog
cross-entropy loss formulation) as our backbone.

\paragraph{Positional encoding in Transformers.}
Self-attention has no intrinsic notion of sequence order: without positional information, it is equivariant to permutations of the input. Therefore, positional and temporal encodings are an important design choice in Transformer-based sequential recommendation. Standard SASRec injects order through learnable absolute positional embeddings, while subsequent Transformer variants explore alternative parameterizations such as relative and rotary positional encodings~\cite{transformer2017,rope2023}. Time-aware recommenders extend this idea by incorporating temporal gaps between interactions. In particular, TiSASRec~\cite{tisasrec} augments self-attention with both absolute position and time-interval information, making it a relevant temporal baseline for our setting.

In parallel, graph Transformers use positional encodings to represent structure rather than sequence order. Laplacian Positional Encoding (LPE), introduced by Dwivedi and Bresson~\cite{Dwivedi2020-GraphTransformer} and later developed in SAN~\cite{Kreuzer2021-SAN} and GraphGPS~\cite{Rampasek2023-GraphGPS}, represents nodes by eigenvectors of the graph Laplacian. These spectral coordinates provide a permutation-aware structural notion of position. Our work adapts this idea to sequential recommendation by using Laplacian eigenvectors of the item co-occurrence graph as graph-derived positional embeddings in SASRec.

\paragraph{Graph-aware sequential recommenders.}

Prior work adds graph-derived or orderless signals to sequential recommenders through extra representation components. MCLSR~\cite{Wang2022-MCLSR} co-trains graph and sequential branches through contrastive learning. MRGSRec~\cite{Baikalov2024-MRGSRec} fuses a sequence encoder with a LightGCN-style graph encoder; GSAU~\cite{gsau} aligns graph and sequential encoders in a shared embedding space; and LOOM~\cite{loom} transfers information from ordered to orderless representations via one-way distillation. These methods show that structural signals can complement sequential modeling, but require extra branches, objectives, or fusion modules. In contrast, we use the item graph only once, before training, to compute spectral item coordinates. The graph signal is then a frozen SASRec lookup in the positional slot, leaving the backbone, loss, and inference path unchanged.

\section{Method}
\label{sec:method}

\subsection{Problem statement}
We consider the standard next-item prediction task. Let $\mathcal{U}$
be a set of users and $\mathcal{I}$ a set of $N$ items. Each user
$u \in \mathcal{U}$ is described by a chronologically ordered
sequence of interactions $S^{u} = (i^{u}_{1}, i^{u}_{2}, \ldots,
i^{u}_{|S^{u}|})$ with $i^{u}_{t} \in \mathcal{I}$. Given the prefix
$S^{u}_{<T}$, the model must rank items in $\mathcal{I}$ by the
predicted probability that the user will interact with each item
next.

We build an undirected weighted item co-occurrence graph $G = (\mathcal{I}, E)$ from the training interactions only. Let $R$ be the binary user-item interaction matrix, where $R_{ui}=1$ if user $u$ interacted with item $i$ in the training split. We construct the item-item adjacency matrix as $A= R^TR$ and set its diagonal to zero. The edge weight $A_{ij}$ counts the number of users whose interaction histories contain both items $i$ and $j$. 

\subsection{Laplacian positional encoding}

Let $D$ be the diagonal degree matrix $D_{ii} = \sum_j A_{ij}$.
We use the symmetric normalized Laplacian \begin{equation} L = I - D^{-1/2} A D^{-1/2}. \end{equation} The normalized Laplacian reduces the effect of item popularity by degree-normalizing the co-occurrence graph.
It admits the eigendecomposition $L = U \Lambda U^{\top}$, where
the columns of $U \in \mathbb{R}^{N \times N}$ are orthonormal
eigenvectors and $\Lambda = \mathrm{diag}(\lambda_{1}, \ldots,
\lambda_{N})$ with $0 = \lambda_{1} \leq \lambda_{2} \leq \cdots
\leq \lambda_{N} \leq 2$. 

The degree normalization reduces the dominance of highly popular items in the co-occurrence graph. We use eigenvectors of $L$ as structural item coordinates. Specifically, we select $k$ non-trivial eigenvectors and stack them into the Laplacian positional encoding matrix \begin{equation} P \in \mathbb{R}^{N \times k}. \end{equation} Each row $P_i$ is a $k$-dimensional graph-spectral representation of item $i$.

When $k=d$, where $d$ is the Transformer hidden size, we use $P$ directly. When $k \neq d$, we apply a learnable projection $W_p \in \mathbb{R}^{k \times d}$: \begin{equation} \mathrm{LPE}[i] = \begin{cases} P_i, & k=d, \\ P_i W_p, & k \neq d. \end{cases} \end{equation} The graph, the normalized Laplacian, and the selected eigenvectors are computed once before training. The resulting LPE table is kept frozen throughout training and inference.

\subsection{Eigenvector selection and model integration}
Our main variant, \textbf{SASRec-LapPE}, uses the eigenvectors corresponding to the $k=d$ smallest non-zero eigenvalues of the normalized Laplacian. This choice captures low-frequency structural variation over the item co-occurrence graph. We also evaluate \textbf{SASRec-LapPE-MLP}, where the same spectral input is passed through a learnable two-layer MLP instead of a linear projection.

LPE is plugged into SASRec by modifying only the input embedding layer. For an item $i_t$ at position $t$ in the input window, the input representation in SASRec-LapPE is \begin{equation} h^{(0)}_{t} = E[i_t] + \alpha \cdot \mathrm{LPE}[i_t], \end{equation} where $E[i_t]$ is the learnable item embedding and $\alpha$ is a learnable scalar controlling the contribution of the graph-spectral signal. This replaces the original SASRec input representation \begin{equation} h^{(0)}_{t} = E[i_t] + p_t, \end{equation} where $p_t$ is the learnable sequence-position embedding.

The LPE vectors are fixed and not updated during training. Unlike the learnable sequence-position embedding $p_t$, $\mathrm{LPE}[i_t]$ depends only on the item identity and the training co-occurrence graph, not on the sequence position $t$, timestamp, or time interval. Thus, SASRec-LapPE keeps chronological order through the ordered sequence and causal mask, but uses no explicit sequence- or time-position embeddings.

\section{Experiments and Results}
\label{sec:experiments}
\subsection{Datasets}
We evaluate on four standard sequential-recommendation benchmarks: Amazon-Beauty, Amazon-Clothing, Amazon-Sports~\cite{amazon_review_data_jmcauley}, and Yelp2018~\cite{lightgcn2020}.
Following common practice~\cite{sasrec2018, bert4rec}, all interactions are
treated as positive feedback, users with fewer than three interactions are removed, and each user's interactions are sorted by timestamp. Dataset statistics after preprocessing are shown in
Table~\ref{tab:dataset_statistics}.
\begin{table}[t]
\centering
\caption{Dataset statistics after preprocessing}
\label{tab:dataset_statistics}
\begin{tabular}{lrrr}
\toprule
Dataset & \#Users & \#Items & \#Interactions \\
\midrule
Amazon-Beauty   & 22 363 & 12 101 & 198 502 \\
Amazon-Clothing & 39 387 & 23 033 & 278 677 \\
Amazon-Sports   & 35 598 & 18 357 & 296 337 \\
Yelp2018 & 31 668 & 38 048 & 1 561 406 \\
\bottomrule
\end{tabular}
\end{table}

\subsection{Experimental setup}
We use a global temporal split evaluation protocol \cite{gusak2025time}. For each user, we split interactions chronologically. The most recent
$20\%$ of interactions is used for testing. From the remaining earlier
interactions, the most recent $10\%$ is used for validation and the
rest for training. For validation, we
use the last validation interaction as the target and the preceding
interactions as the input prefix. For testing, we average results over
ten random holdout splits, where one interaction is sampled from the
test period of each user and all earlier interactions form the prefix. This protocol ensures that validation and test targets always occur after the training interactions in time. The item co-occurrence graph and all LPE vectors are computed from the training interactions only, so no validation or test interactions are used when constructing the graph-derived encodings.

We report ranking metrics NDCG@$k$, Recall@$k$ and diversity metric Coverage@$k$ for $k \in \{10, 100\}$.

We tune hyperparameters with Optuna. The objective is validation NDCG@10. To isolate the positional or structural encoding from backbone re-tuning, we first tune the SASRec backbone for each dataset using 300 trials over learning rate, hidden size, number of heads, dropout, batch size, and
the number of Transformer blocks. We then reuse the best configuration for all SASRec-based variants, except TiSASRec, whose larger memory footprint requires a reduced batch size.

\subsection{Baselines and variants}
Our main baseline is SASRec with learnable sequence-position embeddings, trained with a full-softmax cross-entropy objective rather
than the original sampled binary loss. For each non-padding position,
the model predicts the next item over the full item catalog:
\begin{equation}
\mathcal{L}
=
-\frac{1}{|\mathcal{B}|}
\sum_{(u,t)\in\mathcal{B}}
\log
\frac{\exp(h_t^\top E[y_t])}
{\sum_{j \in \mathcal{I}} \exp(h_t^\top E[j])},
\end{equation}
Here $\mathcal{B}$ denotes the set of all non-padding positions, $h_t$ is the SASRec hidden state at position $t$, $y_t$ is the
target next item, and $E[j]$ is the embedding of item $j$.
We compare SASRec against two LPE variants described in Sec.~\ref{sec:method}: SASRec-LapPE, SASRec-LapPE-MLP. 

We also compare against two alternative positional or temporal
encoding schemes. {SASRec-RoPE} replaces learnable absolute
position embeddings with rotary positional embeddings. {TiSASRec}
extends self-attention with time-interval information. Since TiSASRec
is sensitive to the temporal discretization scale, we perform a grid
search over
$\text{max\_time\_interval} \in \{16, 64, 256, 512, 1024\}$ and report
the configuration with the best validation performance.

For each dataset, we select the best TiSASRec configuration according to validation performance and report the corresponding test results.

All SASRec-based variants share the same sequential backbone. The
difference is where the positional or temporal signal is applied.
SASRec-LapPE computes its graph-spectral signal once offline and uses it
during training and inference as a fixed item-level lookup followed by
a scalar weighting. In contrast, RoPE applies position-dependent
rotations inside the attention layers, while TiSASRec introduces
time-interval information into attention. Thus, SASRec-LapPE adds no
per-layer positional computation and no pairwise temporal encoding
during online training or inference.

\begin{table*}[t]
\centering
\caption{Main results across four sequential-recommendation
benchmarks. Our proposed model, \textbf{SASRec-LapPE}, uses the $d$
smallest non-trivial eigenvectors of the item co-occurrence graph
Laplacian as a frozen positional embedding; \textbf{SASRec-LapPE-MLP}
additionally passes them through a learnable two-layer MLP. We
compare against vanilla SASRec, a positional-embedding-free variant
(\textbf{SASRec-NoPE}), rotary positional embeddings
(\textbf{SASRec-RoPE}), and time-interval self-attention
(\textbf{TiSASRec}). Values are mean (std) in percentage over
ten random holdout splits.}
\label{tab:main_results}
\setlength{\tabcolsep}{6pt}
\renewcommand{\arraystretch}{1.15}
\resizebox{\textwidth}{!}{%
\begin{tabular}{l l c c c c c c}
\toprule
\textbf{Dataset} & \textbf{Metric} & \textbf{SASRec} & \textbf{SASRec-NoPE} & \textbf{SASRec-RoPE} & \textbf{TiSASRec} & \textbf{SASRec-LapPE-MLP} & \textbf{SASRec-LapPE} \\

\specialrule{1.1pt}{4pt}{4pt}

\multirow{6}{*}{\textbf{Beauty}}
& NDCG@10      & 1.73\,(0.09) & 1.74\,(0.06) & 1.68\,(0.08) & \underline{1.75\,(0.06)} & 1.68\,(0.06) & \textbf{1.85\,(0.08)} \\
& NDCG@100     & \textbf{3.89\,(0.06)} & 3.71\,(0.07) & 3.72\,(0.06) & \underline{3.79\,(0.06)} & 3.77\,(0.07) & \textbf{3.89\,(0.08)} \\
& Recall@10    & 3.33\,(0.16) & 3.39\,(0.09) & 3.32\,(0.16) & \underline{3.44\,(0.12)} & 3.20\,(0.08) & \textbf{3.65\,(0.14)} \\
& Recall@100   & \textbf{14.61\,(0.20)} & 13.68\,(0.19) & 14.00\,(0.21) & 14.05\,(0.24) & 14.11\,(0.25) & \underline{14.47\,(0.23)} \\
& Coverage@10  & 64.41\,(0.24) & \textbf{78.13\,(0.28)} & \underline{71.22\,(0.26)} & 65.33\,(0.23) & 41.65\,(0.12) & 67.62\,(0.18) \\
& Coverage@100 & 93.59\,(0.10) & \textbf{96.56\,(0.04)} & \underline{95.94\,(0.06)} & 94.02\,(0.09) & 84.69\,(0.12) & 94.85\,(0.09) \\

\specialrule{1.1pt}{4pt}{4pt}

\multirow{6}{*}{\textbf{Sports}}
& NDCG@10      & 1.29\,(0.08) & 1.18\,(0.06) & \textbf{1.37\,(0.05)} & 0.96\,(0.05) & \underline{1.34\,(0.03)} & 1.33\,(0.05) \\
& NDCG@100     & 2.81\,(0.06) & 2.67\,(0.07) & 2.88\,(0.05) & 2.35\,(0.06) & \textbf{2.95\,(0.05)} & \underline{2.92\,(0.06)} \\
& Recall@10    & 2.42\,(0.15) & 2.22\,(0.09) & \textbf{2.60\,(0.11)} & 1.87\,(0.07) & \underline{2.56\,(0.07)} & 2.53\,(0.08) \\
& Recall@100   & 10.33\,(0.18) & 9.92\,(0.18) & 10.44\,(0.13) & 9.15\,(0.22) & \textbf{10.90\,(0.20)} & \underline{10.86\,(0.16)} \\
& Coverage@10  & \textbf{73.04\,(0.14)} & 52.01\,(0.18) & 59.14\,(0.14) & 4.41\,(0.03) & 52.60\,(0.12) & \underline{62.21\,(0.15)} \\
& Coverage@100 & \textbf{96.25\,(0.06)} & 86.89\,(0.15) & 92.07\,(0.09) & 21.81\,(0.10) & 87.03\,(0.09) & \underline{93.78\,(0.07)} \\

\specialrule{1.1pt}{4pt}{4pt}

\multirow{6}{*}{\textbf{Clothing}}
& NDCG@10      & 0.89\,(0.05) & 0.90\,(0.04) & 0.89\,(0.03) & \underline{1.04\,(0.05)} & 0.93\,(0.05) & \textbf{1.11\,(0.06)} \\
& NDCG@100     & 1.80\,(0.06) & 1.87\,(0.05) & 1.81\,(0.04) & \underline{2.06\,(0.06)} & 1.84\,(0.06) & \textbf{2.29\,(0.05)} \\
& Recall@10    & 1.62\,(0.08) & 1.67\,(0.06) & 1.67\,(0.05) & \underline{1.93\,(0.09)} & 1.72\,(0.08) & \textbf{2.08\,(0.10)} \\
& Recall@100   & 6.37\,(0.14) & 6.74\,(0.13) & 6.44\,(0.13) & \underline{7.21\,(0.16)} & 6.48\,(0.15) & \textbf{8.22\,(0.12)} \\
& Coverage@10  & 59.79\,(0.17) & \textbf{70.99\,(0.21)} & \underline{66.87\,(0.28)} & 52.30\,(0.16) & 60.14\,(0.21) & 48.89\,(0.18) \\
& Coverage@100 & 90.84\,(0.12) & \textbf{96.09\,(0.08)} & \underline{94.88\,(0.08)} & 87.79\,(0.07) & 90.64\,(0.13) & 85.01\,(0.11) \\

\specialrule{1.1pt}{4pt}{4pt}

\multirow{6}{*}{\textbf{Yelp2018}}
& NDCG@10      & 1.89\,(0.09) & \underline{2.41\,(0.15)} & \textbf{2.56\,(0.17)} & 2.21\,(0.13) & 2.35\,(0.17) & 2.38\,(0.09) \\
& NDCG@100     & 4.98\,(0.15) & 5.85\,(0.15) & \textbf{6.08\,(0.17)} & 5.52\,(0.13) & 5.83\,(0.23) & \underline{5.93\,(0.17)} \\
& Recall@10    & 3.84\,(0.18) & \underline{4.85\,(0.33)} & \textbf{5.15\,(0.26)} & 4.52\,(0.21) & 4.73\,(0.27) & 4.75\,(0.13) \\
& Recall@100   & 20.12\,(0.53) & 22.94\,(0.52) & \textbf{23.56\,(0.51)} & 21.95\,(0.35) & 23.03\,(0.63) & \underline{23.37\,(0.70)} \\
& Coverage@10  & 18.86\,(0.26) & 19.59\,(0.13) & \underline{21.46\,(0.14)} & 18.10\,(0.17) & 21.17\,(0.16) & \textbf{23.81\,(0.23)} \\
& Coverage@100 & 58.86\,(0.43) & 61.11\,(0.37) & \underline{63.57\,(0.29)} & 56.25\,(0.31) & 63.52\,(0.27) & \textbf{67.47\,(0.44)} \\

\bottomrule
\end{tabular}%
}
\end{table*}

\begin{figure*}[!t]
  \centering
  \includegraphics[
  width=0.95\textwidth,
  trim=0cm 5cm 0cm 5cm,
  clip
]{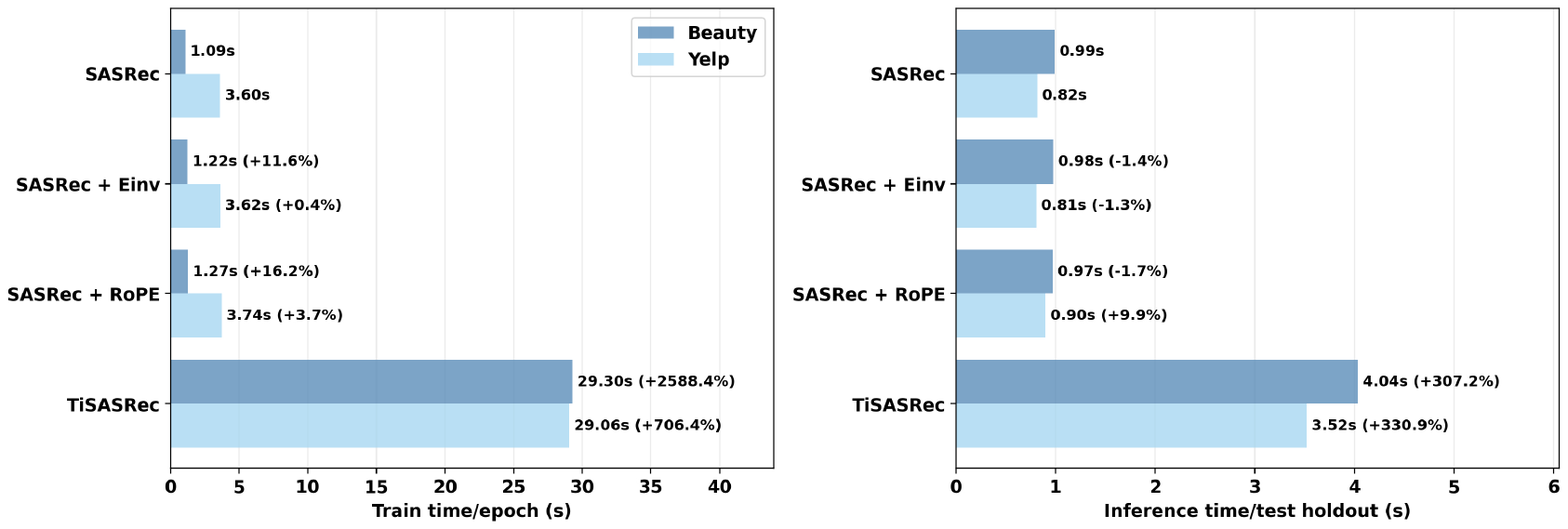}
  \caption{Training time per epoch and inference time per test holdout
  on Beauty and Yelp2018 under identical shared model hyperparameters.
  Percentages are relative to SASRec with its original learnable
  positional embeddings.}
  \Description{Two horizontal bar charts compare SASRec, SASRec with
  Laplacian eigenvector embeddings, SASRec with rotary positional
  embeddings, and TiSASRec. The left chart reports training time per
  epoch on Beauty and Yelp2018; the right chart reports inference time
  per test holdout. TiSASRec is substantially slower than the other
  methods in both charts.}
  \label{fig:efficiency}
\end{figure*}

\subsection{Results}
Results across the four benchmarks appear in Table~\ref{tab:main_results}.
SASRec-LapPE improves over SASRec on most ranking
metrics, with the largest gains on Amazon Clothing and Yelp2018. On
Amazon Clothing, NDCG@10 increases from $0.89$ to $1.11$, NDCG@100
from $1.80$ to $2.29$, and Recall@100 from $6.37$ to $8.22$. On
Yelp2018, SASRec-LapPE improves NDCG@100 from $4.98$ to $5.93$ and
Recall@100 from $20.12$ to $23.37$. These results show that frozen
graph-spectral item encodings provide useful structural information
beyond the learned item embeddings of SASRec.

SASRec-NoPE is an important baseline because it separates the effect
of removing learnable sequence positions from the effect of injecting
graph structure. NoPE is strong on several metrics, especially on
Coverage and on Yelp2018 at cutoff 10. Nevertheless, SASRec-LapPE
improves over SASRec-NoPE on @100 ranking metrics across all four
datasets and on @10 ranking metrics on Amazon-Beauty, Amazon-Sports,
and Amazon-Clothing. This suggests that the gains are not merely due
to deleting $p_t$: replacing it with graph-spectral item structure
provides additional ranking signal.



Compared with stronger positional and temporal baselines, model SASRec-LapPE remains highly competitive while avoiding online positional or temporal computation. The strongest effect appears at the larger cutoff: on the reported @100 ranking metrics, SASRec-LapPE outperforms
TiSASRec across all four datasets and improves over RoPE on three
of the four datasets. Unlike TiSASRec and RoPE, it uses neither timestamps nor position-dependent transformations inside attention layers; its structural signal is computed once from the training item graph and then used as a frozen item-level lookup.

Coverage results further support the usefulness of graph-derived positional information. SASRec-LapPE is consistently stronger than TiSASRec in Coverage@10 and Coverage@100 and improves over SASRec on Amazon-Beauty and Yelp2018. On Yelp2018, it achieves the best Coverage while also improving ranking quality. The main exception is Amazon-Clothing, where large NDCG and Recall gains come with lower Coverage, suggesting that low-frequency Laplacian components may concentrate recommendations within coherent item communities. Overall, Laplacian positional embeddings affect both accuracy and catalog spread, making spectral frequency selection an important direction for future work.

\subsection{Efficiency Comparison}
Unlike the compared positional and temporal alternatives, SASRec-LapPE
introduces additional overhead only before model training, when the
item co-occurrence graph is constructed and its eigendecomposition is
computed to obtain the graph-derived embeddings. In our experiments, we
used an iterative Lanczos solver with approximate time complexity
$O(k|E|+Nk^2)$ and memory complexity $O(|E|+Nk)$, where $N$ and $|E|$
denote the numbers of items and non-zero graph edges, respectively. This
one-time preprocessing, including both graph construction and
eigendecomposition, took $2.52$ seconds on Beauty and $95.86$ seconds on
Yelp2018.

After preprocessing, SASRec-LapPE remains close to SASRec in online
efficiency under identical shared model hyperparameters
(Figure~\ref{fig:efficiency}). Training time is comparable, while
inference is marginally faster. Unlike SASRec, which learns full
positional vectors, our variant freezes the Laplacian embeddings and
learns only the scalar $\alpha$. TiSASRec is substantially slower in
both training and inference.

\section{Conclusion and Future Work}
\label{sec:conclusion}

Our results suggest a simple but useful shift in how positional embeddings are viewed in sequential recommendation. Instead of serving only as markers of interaction order, they can also provide a channel for static structure in the item space. In SASRec, this shift is enough to make frozen Laplacian embeddings from the item co-occurrence graph a viable substitute for standard learnable positional embeddings. The gains over SASRec, together with competitive performance against stronger positional and temporal baselines, indicate that the item graph captures recommendation signal that ordinal encodings do not always recover. This makes graph-derived positional embeddings an attractive lightweight alternative to graph--sequence hybrid models: the graph is used once, before training, while the sequential backbone remains unchanged. An important direction for future work is to understand which parts of the graph spectrum are most useful for recommendation and to design systematic frequency-selection procedures, such as validation-based spectral band search or learnable spectral filtering.

\begin{acks}
The article is an output of a research project HSE-BR-2025-019
implemented as part of the Basic Research Programm at HSE University.
This research was supported in part through computational resources of
HPC facilities at HSE University~\cite{kostenetskiy2021hpc}.
\end{acks}

\section*{GenAI Usage Disclosure}
This work complies with the CIKM 2026 GenAI usage policy. The
authors disclose the following use of a generative AI assistant
(Anthropic Claude) during manuscript preparation.

\textbf{Writing.} The narrative, the framing of the contribution,
and the initial idea-level drafts of every section of the manuscript
were written by the authors. The AI assistant was used to scaffold the overall paper
structure, to
suggest LaTeX boilerplate, and to check for errors and verify the
consistency of data presented in the prose against the numbers
reported in the results tables. All AI-generated text was reviewed,
edited, and verified by the authors prior to inclusion.
The authors take full responsibility for the final wording.

\textbf{Code.} AI tools were used to assist with scripts and optimization of implementation. The authors reviewed, modified and tested all generated code. 

\textbf{Data.} No GenAI tools were used to generate or augment the
data used in this work. All datasets are publicly available and
obtained from their original sources as cited in
Section~\ref{sec:experiments}.

\textbf{Experiments and analysis.} All experimental results,
hyperparameter searches, and statistical aggregations were produced
by the authors. The AI assistant was used to help describe these
results during manuscript writing, but the experimental design,
execution, and analytical conclusions are the authors'.

The intellectual contributions of this work --- the research
question, the proposed method, the experimental design, and the
conclusions --- are the authors' own.

\bibliographystyle{ACM-Reference-Format}
\bibliography{refs}

\end{document}